\documentclass[]{spie}  %>>> use for US letter paper
\usepackage[]{graphicx}

\newcommand{\bp}{{\bf p}}
\newcommand{\br}{{\bf r}}

\newcommand{\bj}{{\bf j}}
\newcommand{\bE}{{\bf E}}
\newcommand{\eps}{\varepsilon}
\newcommand{\la}{\langle}
\newcommand{\ra}{\rangle}

\newcommand{\qw}{q_{\omega}}
\newcommand{\dx}{\delta x}
\newcommand{\bx}{\bar{x}}
\newcommand{\Kt}[3]{ K(#2-#3) }
\newcommand{\xx}{\la \dx(t_1)\dx(t_2) \ra}
\newcommand{\xxx}{\la \dx(t_1)\dx(t_2)\dx(t_3) \ra}
\newcommand{\dQ}{ \left.\frac{dQ}{dx}\right|_{\bx} }
\newcommand{\ddQ}{ \left.\frac{d^2Q}{dx^2}\right|_{\bx} }
\renewcommand{\varphi}{\phi}

\title{Semiclassical noise beyond the second cumulant}

\author{Kirill E. Nagaev
\skiplinehalf
William I. Fine Theoretical Physics Institute,
University of Minnesota,
116 Church Street S.E.
Minneapolis, MN 55455
USA
}

\authorinfo{ E-mail: nagaev@physics.umn.edu}

\begin{document}
  \maketitle

%%%%%%%%%%%%%%%%%%%%%%%%%%%%%%%%%%%%%%%%%%%%%%%%%%%%%%%%%%%%%
\begin{abstract}
We show how the semiclassical Langevin method can be extended to calculations
of higher-than-second cumulants of noise. These cumulants are affected by
indirect correlations between the fluctuations, which may be considered as
"noise of noise." We formulate simple diagrammatic rules for calculating the
higher cumulants and apply them to mesoscopic diffusive contacts and chaotic
cavities. As one of the application of the method, we analyze the frequency
dependence of the third cumulant of current in these systems and show that it
contains additional peculiarities as compared to the second cumulant. The
effects of environmental feedback in measurements of the third cumulant are
also discussed in terms of this method.
\end{abstract}


%%%%%%%%%%%%%%%%%%%%%%%%%%%%%%%%%%%%%%%%%%%%%%%%%%%%%%%%%%%%%
\section{INTRODUCTION}
\label{sect:intro}

Since the nonequilibrium noise in mesoscopic systems became a hot topic in late
80s, the semiclassical Boltzmann--Langevin approach has been successfully
competing with quantum-mechanical methods in these studies.\cite{Blanter-00}
While restricted to certain class of systems and parameters, this approach is
generally more simple and physically appealing. It can easily handle some of
the problems that are extremely difficult to solve by directly using
quantum-mechanical equations.

Higher-order correlations of current became a subject of interest for
theorists since early nineties. This work was pioneered by Levitov and Lesovik
\cite{Levitov-93}, who discovered that the charge transmitted through a
single-channel quantum contact obeys a binomial distribution. Based on their
quantum-mechanical formulas, higher cumulants of current were calculated for a
variety of multichannel phase-coherent systems
\cite{Jalabert-94,Baranger-94,Lee-95}. Meanwhile a wide class of these systems,
like diffusive conductors and chaotic cavities, allows a semiclassical
description of their average transport properties and second cumulant of noise
\cite{Kogan-69,Gantsevich-69}. Hence it is of interest to have a fully
semiclassical theory for higher cumulants of noise in these systems.

Below we present an extension of the Boltzmann--Langevin approach that allows
one to calculate the statistical properties of noise beyond the second
cumulant. This extension is based (a) on the
time-scale separation between the correlation time of microscopic sources
of noise acting upon the system ("noise generators") and the response of the
system to them and (b) on the smallness of fluctuations. It has been initially
proposed for the higher cumulants of current in mesoscopic diffusive
wires,\cite{Nagaev-02a} where the resulting terms could be directly mapped
on the quantum-mechanical diagrams for these quantities.\cite{Gutman-03}
Shortly after this, the cascade expansion was postulated for a chaotic
cavity with quantum contacts and tested against quantum-mechanical results
for the third and fourth cumulants of current.\cite{Nagaev-02b} Very
recently, a general proof of this expansion was given for semiclassical
cumulants of arbitrary order.\cite{Jordan-04,Gutman-04}

An application of the semiclassical method to calculations of the frequency
dependence of the third cumulant of current noise in mesoscopic diffusive
wires\cite{Pilgram-04} and chaotic cavities\cite{Nagaev-03} gave very
unexpected results. Unlike the conventional noise and ac electric response that
exhibit a dispersion only at the inverse $RC$ time of the system, the third
cumulant of current has also a dispersion at the inverse dwell time of an
electron on the system.  This dispersion is due to slow fluctuations of the
distribution function that do not violate electroneutrality and are akin to
fluctuations of local temperature. These fluctuations do not directly
contribute to the current and therefore are not seen in conventional noise, but
they modulate the intensity of noise sources and therefore manifest themselves
in higher correlations of current. As the dwell time is generally much larger
than the $RC$ time, the resulting dispersion may be experimentally observed.

The cascaded expansion of higher cumulants is valid not only for systems
described by Boltzmann equation but also for other systems that satisfy the
conditions of time-scale separation and smallness of fluctuations. Recently,
this method has been used for calculating the feedback of the external circuit
in measurements of the third cumulant of an arbitrary mesoscopic
resistor.\cite{Beenakker-03}

The paper is organized as follows. In Section~\ref{sect:general} we describe the
general cascaded formalism. We give a brief overview of the standard
Boltzmann--Langevin method, then we derive the expression for the third
cumulant using a generalized Fokker--Planck equation. After this, we formulate
the general diagrammatic rules for calculating cumulants of arbitrary order. In
Section~\ref{sect:appl} we calculate the frequency dependence of the third
cumulant for a mesoscopic diffusive wire and a chaotic cavity. Finally, we
address the effect of the external circuit on the measurements of the third
cumulant. The main conclusions are summarized in Section~\ref{summary}.

\section{GENERAL FORMALISM}
\label{sect:general}

The basis for the semiclassical description of kinetics is the existence of
two well separated time scales, one of which describes a "slow" classical
evolution of the system and the other describes "fast" quantum processes.
For example, the collision integral in the Boltzmann equation may be
written as local in time because quantum-mechanical scattering is assumed
to be fast as compared to the evolution of the distribution function. For
this reason, it is possible to describe the noise associated with electron
scattering by  Langevin sources that are $\delta$-correlated in time. On
the other hand, using the semiclassical kinetic equation implies
coarse-grain averaging over $N = (\Delta p \Delta x/2\pi\hbar)^3$
states located in an elementary cell of phase space whose characteristic
dimensions $\Delta p$ and $\Delta x$ are small as compared to the
characteristic scales of the problem. This suggests that the cumulants of the
distribution function $f$ should decay with number as $1/N^{n-1}$.

\subsection{Boltzmann--Langevin Method}
\label{sect:BL}

The second cumulants of noise are conveniently described  by the
Boltzmann--Langevin equation, \cite{Kogan-69}  which is obtained by
linearizing the standard Boltzmann equation with respect to a fluctuation of
the distribution function  $\delta f$. It reads
\begin{equation}
  \left[
     \frac{\partial}{\partial t}
     +
     {\bf v}\frac{\partial}{\partial{\bf r}}
     +
     e{\bf Ev}\frac{\partial}{\partial \eps}
  \right] \,
  \delta f(\bp,\br,t)
  +
  \delta I
  =
  -e\,\delta{\bf E\,v} \frac{\partial f}{\partial\eps}
  +
  J^{ext},
  \label{BL}
  \end{equation}
where $\delta I$ is the linearized collision integral and $J^{ext}$ is the
Langevin source that accounts for the randomness of electron scattering.
This equation should be supplemented by a self-consistency equation for
the electric field
\begin{equation}
\nabla\delta\bE
  =
  4\pi\delta\rho,
\quad
  \delta\rho(\br,t)
  =
  e\sum\limits_{\bp}
  \delta f(\bp, \br, t).
\label{self-cons}
\end{equation}
The key issue in the Boltzmann--Langevin method is a derivation of the
correlation function of Langevin sources. Several decades ago,Kogan and
Shulman\cite{Kogan-69} proposed a physically appealing expression fort this
function. Because of a short duration of scattering events, the Langevin
sources are $\delta$-correlated in time. Furthermore, since different
scattering events are independent, the scattering between each pair of states
$(\bp,\br)$ and $(\bp',\br)$ presents a Poissonian process whose cumulants of
any order are proportional to its average rate
\begin{equation}
 J(\bp\to\bp')
 =
 W(\bp, \bp')
 f(\bp,\br, t)[1 - f(\bp', \br, t)].
\label{flux}
\end{equation}
Hence the cumulants  of extraneous sources of the corresponding order related
with randomness of scattering may be written as the sums of incoming and
outgoing scattering fluxes taken with appropriate signs. For example, the
second cumulant is given by\cite{Kogan-69}
$$
 \la\la
   J^{ext}(\bp_1,\br_1,t_1)
   J^{ext}(\bp_2,\br_2,t_2)
 \ra\ra
 =
 \delta(\br_1 - \br_2)
 \delta(t_1 - t_2)
 \Biggl\{
   \delta_{\bp_1\bp_2}
   \sum\limits_{\bp'}
   [
    J(\bp_1 \rightarrow \bp' )
    +
    J(\bp'  \rightarrow \bp_1)
   ]
$$ \begin{equation}
   -
   J(\bp_1 \rightarrow \bp_2)
   -
   J(\bp_2 \rightarrow \bp_1)
 \Biggr\}.
 \label{JJ}
\end{equation}
To calculate, e.g., the correlator of  current fluctuations $\la\delta
j(\br_1,t_1)\delta j(\br_2,t_2)\ra$, one has to take two formal solutions of
Eq. (\ref{BL}) in $\delta f$, calculate the two fluctuations of current via
\begin{equation}
 \delta\bj(\br,t)
 =
 e\sum\limits_{\bp}
 {\bf v}\delta f(\bp,\br,t)
 \label{dj},
\end{equation}
multiply them, and average the product using Eq.(\ref{JJ}). As the fluxes
(\ref{flux}) depend on the average distribution function,the resulting noise
is also a functional of $f$.

The Boltzmann--Langevin method has been very successful in predicting the $1/3$
suppression of shot noise in mesoscopic diffusive contacts\cite{Nagaev-92} and
analyzing the effects of electron-electron scattering on it.\cite{Nagaev-95}

\subsection{Calculation of the Third Cumulant}
\label{sect:third}

One might think that all higher cumulants of order $n$ could be treated
similarly just by averaging the product of $n$ formal solutions of Eq.
(\ref{BL}) with the corresponding analog of Eq. (\ref{JJ}). However this is
not the case. The problem is that the correlator of Langevin sources (\ref{JJ})
as well as higher-order correlators of $\delta J^{ext}$ are themselves
functions of $f$, which fluctuates in time, and this results in additional
correlations between $\delta f$'s. To elucidate this point, we consider a
simple example of a Markov process where a random quantity $x(t)$ is described
by a Langevin equation
\begin{equation}
 \dot x = -\frac{1}{\tau}(x - \bx) + \xi(t)
 \label{Langevin}
\end{equation}
and $\xi(t)$ is a random extraneous force with zero mean $\la \xi(t) \ra=0$.
Suppose that an $n$th cumulant of $\xi$ is inversely proportional to a large
number $N^{n-1}$ and explicitly depends on $x$:
\begin{equation}
 \la
  \xi(t_1)\xi(t_2)
 \ra
 =
 \frac{1}{N} \delta(t_1 - t_2) Q(x_1),
\qquad
 \la
  \xi(t_1)\xi(t_2) \xi(t_3)
 \ra
 =
 \frac{1}{N^2} \delta(t_1 - t_2) \delta(t_2 - t_3) R(x_1).
 \label{xi3}
\end{equation}
This suggests that as $\Delta t \to 0$, the cumulants of increment of
$x$  are proportional to $\Delta t$ and equal
$$
 \la x(t + \Delta t) - x(t) \ra
 =
 -\frac{\Delta t}{\tau}[x(t) - \bx],
\qquad
 \la[x(t + \Delta t) - x(t)]^2\ra
 =
 \frac{\Delta t}{N} Q(x(t)),
\qquad
\la[x(t + \Delta t) - x(t)]^3\ra
 =
 \frac{\Delta t}{N^2} R(x(t)).
$$
It's a textbook knowledge that the probability $w(x,t)$ in this case obeys a
generalized Fokker--Planck equation in the form
\begin{equation}
 \frac{\partial w}{\partial t}
 =
 \frac{1}{\tau}
 \frac{\partial}{\partial x}
 [(x -\bx)w]
 +
 \frac{1}{2N}
 \frac{\partial^2(Qw)}{\partial x^2}
 -
 \frac{1}{6N^2}
 \frac{\partial^3(Rw)}{\partial x^3}.
 \label{FP}
\end{equation}
By multiplying both parts of Eq. \ref{FP} by $\dx \equiv x(t) - \bx$, $\dx^2$,
and $\dx^3$ and integrating them over $x$ by parts the corresponding number of
times, one obtains equations of motion for the conditional cumulants of $x$
\begin{equation}
 \left(
  \frac{\partial}{\partial x}
  +
  \frac{1}{\tau}
 \right)
 \la\dx(t)\ldots\ra
 =
 0,
%\label{dx}
%\end{equation}
%
\qquad
%\begin{equation}
 \left(
  \frac{\partial}{\partial x}
  +
  \frac{2}{\tau}
 \right)
 \la\dx^2(t)\ldots\ra
 =
 \frac{1}{N}
 \la Q(t) \ra,
 \label{dx^2}
\end{equation}
and
\begin{equation}
 \left(
  \frac{\partial}{\partial x}
  +
  \frac{3}{\tau}
 \right)
 \la\dx^3(t)\ldots\ra
 =
 \frac{3}{N}
 \la \dx(t) Q(t) \ra
 +
 \frac{1}{N^2}
 \la R(t) \ra,
 \label{dx^3}
\end{equation}
where $\ldots$ stand for any functions of $x$ taken at previous instants of
time, $\la Q(t) \ra \equiv \la Q(\bx + \dx(t)) \ra$, and $\la R(t) \ra \equiv
\la R(\bx + \dx(t)) \ra$. By subsequently solving the equations for lower
cumulants with senior cumulants as the initial conditions, one obtains the
expressions for the two- and three-time correlation functions of $\dx$ for
$t_1 > t_2 > t_3$ in the form
\begin{equation}
 \xx
 =
 \frac{1}{N}
 \int\limits_{-\infty}^t dt'\,
 K(t_1 - t') K(t_2 - t') \la Q(t') \ra
 \label{xx-sol2}
\end{equation}
and
$$
 \xxx
 =
 \frac{1}{N}
 \int\limits_{t_3}^{t_2} dt'\,
 K(t_1 - t') K(t_2 - t') \la Q(t')\dx(t_3)\ra
$$ \begin{equation}
 +
 \int\limits_{-\infty}^{t_3} dt'\,
 K(t_1 - t') K(t_2 - t') K(t_3 - t')
 \left[
     \frac{3}{N} \la Q(t')\dx(t')\ra
     +
     \frac{1}{N^2} \la R(t')\ra
 \right],
 \label{xxx-sol2}
\end{equation}
where $K(t-t') = \exp[-(t - t')/\tau]$.

Expand now $Q(t')$ and $R(t')$ in powers of $\dx(t')$ about $\bx$. By
substituting these expansions
$$
 Q(t)
 =
 Q(\bx)
 +
 \dx(t') \left.\frac{dQ}{dx}\right|_{\bx}
 +
 \frac{1}{2}\dx^2(t') \left.\frac{d^2Q}{dx^2}\right|_{\bx}
 +
 \ldots,
 \qquad
 R(t)
 =
 R(\bx)
 +
 \dx(t') \dQ
 +
 \frac{1}{2}\dx^2(t') \ddQ
 +
 \ldots
$$
into Eqs. (\ref{xx-sol2}) and (\ref{xxx-sol2}), one obtains a system of
equations where cumulants of all orders are coupled together and which cannot
be solved in a general case. However one can make use of the smallness of
$1/N$. Retaining only leading terms of the order of $1/N^{n-1}$ in $n$th
cumulants, it is possible to solve the equations recursively by expressing the
higher cumulants in terms of the lower ones. Thus the expression for the second
cumulant becomes just
\begin{equation}
 \xx
 =
 \frac{1}{N}
 \int\limits_{-\infty}^{t_2} dt'\,
 K(t_1 - t') K(t_2 - t') Q(\bx).
 \label{xx-final}
\end{equation}
This is exactly what the standard Langevin approach with $Q=Q(\bx)$ gives.

The expression for the third cumulant appears to be more involved. Taking into
account that
$
 K(t_i - t')
 \la Q(t')\dx(t')\ra
 =
 \la\dx(t_i)Q(t')\ra
$
if $t_i > t'$, it is easily brought to a form
$$
 \xxx
 =
 \frac{1}{N}
 \int\limits_{-\infty}^{t_2} dt'\,
 \Kt{}{t_1}{t'} \Kt{}{t_2}{t'} \dQ \la\dx(t')\dx(t_3)\ra
$$ $$
 +
 \frac{1}{N}
 \int\limits_{-\infty}^{t_3} dt'\,
 \Kt{}{t_2}{t'} \Kt{}{t_3}{t'} \dQ \la\dx(t')\dx(t_1)\ra
%$$ $$
 +
 \frac{1}{N}
 \int\limits_{-\infty}^{t_2} dt'\,
 \Kt{}{t_1}{t'} \Kt{}{t_3}{t'} \dQ \la\dx(t')\dx(t_2)\ra
$$ \begin{equation}
 +
 \frac{1}{N^2}
 \int\limits_{-\infty}^{t_3} dt'\,
 \Kt{}{t_1}{t'} \Kt{}{t_2}{t'}  \Kt{}{t_3}{t'}  R(\bx).
 \label{xxx-der}
\end{equation}
We note that the first three terms of Eq. \ref{xxx-der} present convolutions of
the functional derivatives of the second cumulant with another second cumulant,
so that symbolically, this equation may be recast in a form
\begin{equation}
 \la\dx_1\dx_2\dx_3\ra
 =
 \la\dx_1\dx_2\dx_3\ra_{Q=0}
 +
 P_{123}
 \left\{
  \frac{ \delta\la\dx_1\dx_2\ra }{\delta\bx} \la\delta\bx\dx_3\ra
 \right\},
 \label{xxx-symb}
\end{equation}
where $P_{123}$ denotes a summation over all inequivalent permutations of
indices $(123)$ and $\delta\la\ldots\ra/\delta\bx$ denotes a functional
derivative with respect to $\bx$. The first term in Eq. (\ref{xxx-symb})
presents a direct contribution to the third cumulant of $x$ from the third
cumulant of extraneous sources, and the second term presents the cascade
corrections, which result from modulation of the second cumulant by fluctuations
of $x$.

If $x$ denotes a semiclassical distribution function $f(\bp,\br,t)$, one may
roughly imagine these cascade corrections as fluctuations of Nyquist noise of a
resistor caused by fluctuations of its temperature. It should be noted that the
conditions for validity of the cascaded approach are the same as the conditions
of validity of Boltzmann description itself and no additional assumptions are
needed.

\subsection{Diagrammatic Expansions for Higher Cumulants}
\label{higher}

Similar expressions are valid for higher cumulants. Apart from direct products
of $n$ formal solutions of the Langevin equation, $n$th cumulant also contains
cascade corrections from an interplay between lower cumulants. The cascade
corrections are conveniently presented in a diagrammatic form (see Figs.
{\ref{diag3} and \ref{diag4}). All diagrams present graphs whose outer vertices
correspond to different instances of $x$ and whose inner vertices correspond
either to cumulants of extraneous currents or their functional derivatives. The
number of arrows outgoing from an inner vertex corresponds to the order of the
cumulant and the number of incoming arrows corresponds to the order of a
functional derivative. Apparently, the difference between the total order of
cumulants involved and the total number of functional differentiations should
be equal to the order of the cumulant being calculated. As there should be no
backaction of higher cumulants on lower cumulants, all diagrams are singly
connected. Therefore any diagram for the $n$th cumulant of the current may be
obtained from a diagram of order $m < n$ by combining it with a diagram of
order $n-m+1$, i.e., by inserting one of its outer vertices into one of the
inner vertices of the latter. Hence the most convenient way to draw diagrams
for a cumulant of a given order is to start with diagrams of lower order and to
consider all their inequivalent combinations that give diagrams of the desired
order. The analytical expressions corresponding to each diagram contain
numerical prefactors equal to the numbers of inequivalent permutations of the
outer vertices.

At first, these rules had been postulated,\cite{Nagaev-02a,Nagaev-02b} but very
recently they were proved independently by Jordan et al.\cite{Jordan-04} using
the stochastic path integral approach and by Gutman et al.\cite{Gutman-04}
using the supersymmetry method.
   \begin{figure}
   \begin{center}
   \begin{tabular}{c}
   \includegraphics[height=4cm]{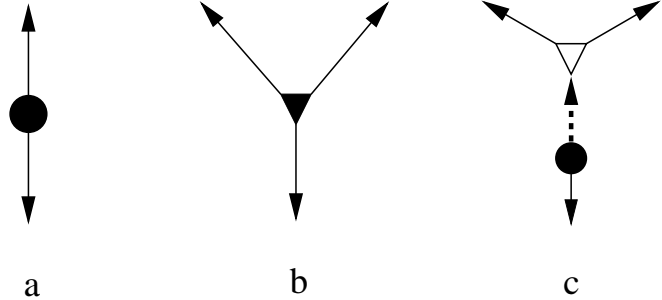}
   \end{tabular}
   \end{center}
   \caption[example]
   { \label{diag3}
   The second cumulant and the two contributions to the third cumulant of
   current. The external ends correspond to current fluctuations at different
   moments of time and the dashed lines, to fluctuations of the distribution
   function. The full circle and triangle correspond to cumulants of
   extraneous currents and the empty triangle to the functional derivative of
   the second cumulant. }
   \end{figure}
   \begin{figure}
   \begin{center}
   \begin{tabular}{c}
   \includegraphics[height=7cm]{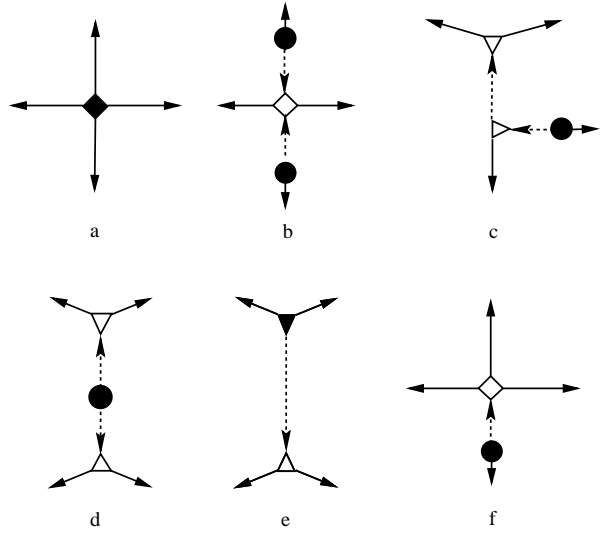}
   \end{tabular}
   \end{center}
   \caption[example]
   { \label{diag4}
   The contributions to the fourth cumulant of the current. Dashed lines
   correspond to fluctuations of the distribution function. Full circles,
   triangles and squares correspond to the second, third, and fourth
   cumulants of extraneous currents. The empty triangles and squares present
   their functional derivatives. }
   \end{figure}

\section{APPLICATIONS OF THE METHOD}
\label{sect:appl}

We now concentrate on the two most important mesoscopic systems that allow a
semiclassical treatment, i.e. diffusive wires and classical chaotic cavities.
Both these types of multichannel systems have a complicated internal dynamics
characterized by essentially different time scales. In the case of a diffusive
wire, thire is a large separation between the elastic scattering time and the
time of diffusion of an electron across the wire. In the case of a chaotic
cavity, there is a large separation between the time of flight through the
cavity and the dwell time of the electron on it. The fluctuations in these
systems are relatively small because of a large number of quantum channels in
them, so the cascaded Langevin approach may be conveniently applied to them.

Of a primary interest to us will be the frequency dependence of the third
cumulant of current
\begin{equation}
 P_3(\omega_1,\omega_2)
 =
 \int d(t_1 - t_2) \int d(t_2 - t_3)
 \exp
 [
  i\omega_1(t_1 - t_3)
  +
  i\omega_2(t_2 - t_3)
 ]
 \la
  \delta I(t_1) \delta I(t_2) \delta I(t_3)
 \ra
 \label{def3}
\end{equation}
in these systems. The point is that in the case of a good
conductor with a short screening length, the dwell time of an electron in the
system is seen neither in its linear response nor in the second cumulant of
current because the charge transport is primarily controlled by the screening
effects whose characteristic time scale is the $RC$ time of the system.
Typically, this time is so short that the resulting dispersion of noise cannot
be experimentally observed. In contrast to this, the third cumulant of current
is affected not only by fluctuations of the total electron number at a given
point, but also by fluctuations of their distribution in energy, which may
affect the intensity of noise sources. The latter fluctuations have a much
longer relaxation time since they do not violate the electroneutrality and
therefore may result in a dispersion at experimentally accessible frequencies.

\subsection{Diffusive wires}
\label{sect:diffusive}

Consider a quasi-one-dimensional diffusive wire of length $L$ and conductivity
$\sigma$ with a constant voltage drop $V$ across it. All dimensions are assumed
to be much larger than the elastic mean free path and the screening length in
the metal. We also assume that dimensions are much smaller than the
electron-phonon scattering length, but take into account electron-electron
scattering. It is assumed that the contact is located close to an external
gate, so it is described by a single capacitance. In the case of a diffusive
metal, the distribution function $f(\eps,\br)$ and its fluctuations are almost
isotropic in the momentum space, and the Boltzmann--Langevin equation reduces
to a stochastic diffusion equation
\begin{equation}
 \left(
  \frac{\partial}{\partial t}
  -
  D\nabla^2
 \right)
 \delta f
 -
 \delta I_{ee}
 =
 -e
 \delta\dot{\phi}
 \frac{\partial f}{\partial\eps}
 -
 \nabla\delta{\bf F}^{imp}
 -
 \delta F^{ee},
 \label{diffusion}
\end{equation}
where $D$ is the diffusion coefficient, $\delta I_{ee}$ is the
linearized electron-electron collision integral, and $\delta{\bf
F}^{imp}$ and $\delta F^{ee}$ are random extraneous sources
associated with electron-impurity and electron-electron
scattering. The fluctuation of the electric potential $\delta\phi$ that
appears in this equation should be calculated self-consistently
from the Poisson equation
\begin{equation}
 \nabla^2\delta\phi
 =
 -4\pi\delta\rho,
\label{Poisson}
\end{equation}
where the fluctuation of charge density $\delta\rho$ is given by
\begin{equation}
 \delta\rho
 =
 eN_F
 \left(
  \int d\eps \delta f(\eps)
  +
  e\delta\phi
 \right)
\label{self-consist}
\end{equation}
and where $N_F$ is the Fermi density of states. In the case of a
quasi-one-dimensional contact, a solution of Eqs. (\ref{diffusion}) -
(\ref{self-consist}) is of the form\cite{Nagaev-98}
\begin{equation}
 \delta\phi(x,\omega)
 =
 \frac{1}{S_0 \sigma}
 \left(
  \nabla^2
  +
  i\omega RC/L^2
 \right)^{-1}
 \frac{\partial}{\partial x}
 \int d^2r_{\perp}
 \delta j_x^{ext}(\br),
 \label{dphi}
\end{equation}
where $x$ is the coordinate along the contact, $S_0$ is the cross section area of the
contact, $C$ and $R$ are the
capacitance and the resistance of the contact, and
\begin{equation}
  \delta\bj^{ext}
  =
  eN_F
  \int d\eps
  \delta{\bf F}^{imp}.
  \label{dj-vs-df}
\end{equation}
A fluctuation of the total current at the left end of the contact equals
\begin{equation}
 \delta I
 =
 \sigma
 \int d^2 r_{\perp}
 \left.
  \frac
  { \partial\delta\phi(x, \omega) }
  { \partial x }
 \right|_{x=-L/2}.
 \label{dI_L}
\end{equation}
Making use of the correlation function of extraneous sources
\begin{equation}
 \la
   \delta F_{\alpha}^{imp}(\eps, \br)
   \delta F_{\beta }^{imp}(\eps', \br')
 \ra_{\omega}
 =
 2\frac{D}{N_F}
 \delta(\br - \br')
 \delta(\eps - \eps')
 \delta_{\alpha\beta}
 f(\eps,\br)[1 - f(\eps,\br)],
 \label{<dF^2>}
\end{equation}
one easily obtains the second-order correlation function for the fluctuations
of the current as a functional of the distribution function
$f$.\cite{Nagaev-92} At frequencies much smaller than $(RC)^{-1}$ it is of the form
\begin{eqnarray}
 \la
  \delta I(\omega_1)
  \delta I(\omega_2)
 \ra
 =
 4\pi
 \delta( \omega_1 + \omega_2 )
 (RL)^{-1}
 \int dx \int d\eps
 f(\eps, x)[1 - f(\eps,x)].
\label{<I^2>}
\end{eqnarray}
%
%   \begin{figure}
%   \begin{center}
%   \begin{tabular}{c}
%   \includegraphics[height=5cm]{wire.eps}
%   \end{tabular}
%   \end{center}
%   \caption[example]
%   { \label{wire}
%   A diffusive wire with an external gate. }
%   \end{figure}
%

To calculate the third cumulant of current in a diffusive wire, one may use the
cascaded Langevin approach described in the previous section with $f$ in place
of $x$. In the case of diffusive metal, the direct contributions from the
higher-than-second cumulants of Langevin sources to the corresponding cumulants
of current are negligibly small because these cumulants are proportional to the
inverse elastic scattering time $\tau^{-1}$ (see Section \ref{sect:BL}) and
each solution of the Boltzmann--Langevin equation for the anisotropic part of
$\delta f$ gives an additional factor of $\tau$, so that the $n$th cumulant of
transport current would be proportional to $\tau^{n-1}$ instead of $\tau$.
Hence all higher cumulants of current for diffusive contacts are dominated by
diagrams constructed of the second cumulant of Langevin sources and its
functional derivatives. As has been shown in Section \ref{sect:third}, the
third cumulant in this case is of the form
\begin{equation}
  \la
    \delta I(t_1)
    \delta I(t_2)
    \delta I(t_3)
  \ra
  =
  P_{123}
  \left\{
   \int dt \int d\eps \int d^3r\,
   \frac{
     \delta
     \la
       \delta I(t_1)
       \delta I(t_2)
     \ra
   }{
     \delta f(\eps, \br, t)
   }
   \la
    \delta f(\eps, \br, t)
    \delta I(t_3)
   \ra
  \right\}.
  \label{I^3-gen}
\end{equation}

We restrict ourselves to the two limiting cases of absolutely elastic
scattering and of a strong electron-electron scattering. In the case of purely
elastic scattering, $\delta I^{ee}$ and $\delta f^{ee}$ in Eq.
(\ref{diffusion}) may be dropped. Furthermore, we consider frequencies much
smaller than $(RC)^{-1}$, so the frequency-dependent term in Eq. (\ref{dphi})
may be neglected and the solution of these equations is straightforward. Using
the well-known expression for the average distribution function
\begin{equation}
 \bar{f}(\eps,x)
 =
 \left(
  \frac{1}{2}
  +
  \frac{x}{L}
 \right)
 f_0(\eps + eV/2)
 +
 \left(
  \frac{1}{2}
  -
  \frac{x}{L}
 \right)
 f_0(\eps - eV/2),
\label{f-elast}
\end{equation}
where $f_0$ is the equilibrium Fermi distribution, one obtains that in the
frequency range of interest,
\begin{equation}
  P_3(\omega_1,\omega_2)
  =
  P(\omega_1) + P(\omega_2) + P(-\omega_1 - \omega_2),
  \label{P}
\end{equation}
where
\begin{eqnarray}
 P_{el}(\omega)
 =
 -\frac{4}{3}
 \frac{e^2V}{R}
 \frac{
  \qw L
  (\qw^2L^2 + 30)
  \sinh(\qw L)
  -
  8
  (\qw^2L^2 + 6)
  \cosh(\qw L)
  +
  2\qw^2 L^2
  +
  48
 }{
  \qw^5 L^5
  \sinh(\qw L)
 }
\label{P-zT}
\end{eqnarray}
at $eV \gg T$ and
\begin{equation}
 P_{el}(\omega)
 =
 \frac{4}{3}
 \frac{e^2V}{R}
 \frac{
  2
  \cosh(\qw L)
  -
  \qw L
  \sinh(\qw L)
  -
  2
 }{
  \qw^3 L^3
  \sinh(\qw L)
 }.
\label{P-zV}
\end{equation}
at $T \gg eV$. Here $q_{\omega}=(i\omega/D)^{1/2}$. At $\omega = 0$ these
expressions give $-(1/45)e^2V/R$ and $-(1/9)e^2V/R$, which corresponds to
$P_3(0,0) = -(1/15)e^2V/R$ and $P_3(0,0) = -(1/3)e^2V/R$. These zero-frequency
results are in agreement with the results of Gutman and Gefen.\cite{Gutman-03}
At finite frequency, Eqs. (\ref{P-zT}) and (\ref{P-zV}) become complex-valued
and tend to zero as $i/\omega$ at $\omega\to\infty$.

\begin{figure}
   \begin{center}
   \begin{tabular}{c}
   \includegraphics[height=6cm]{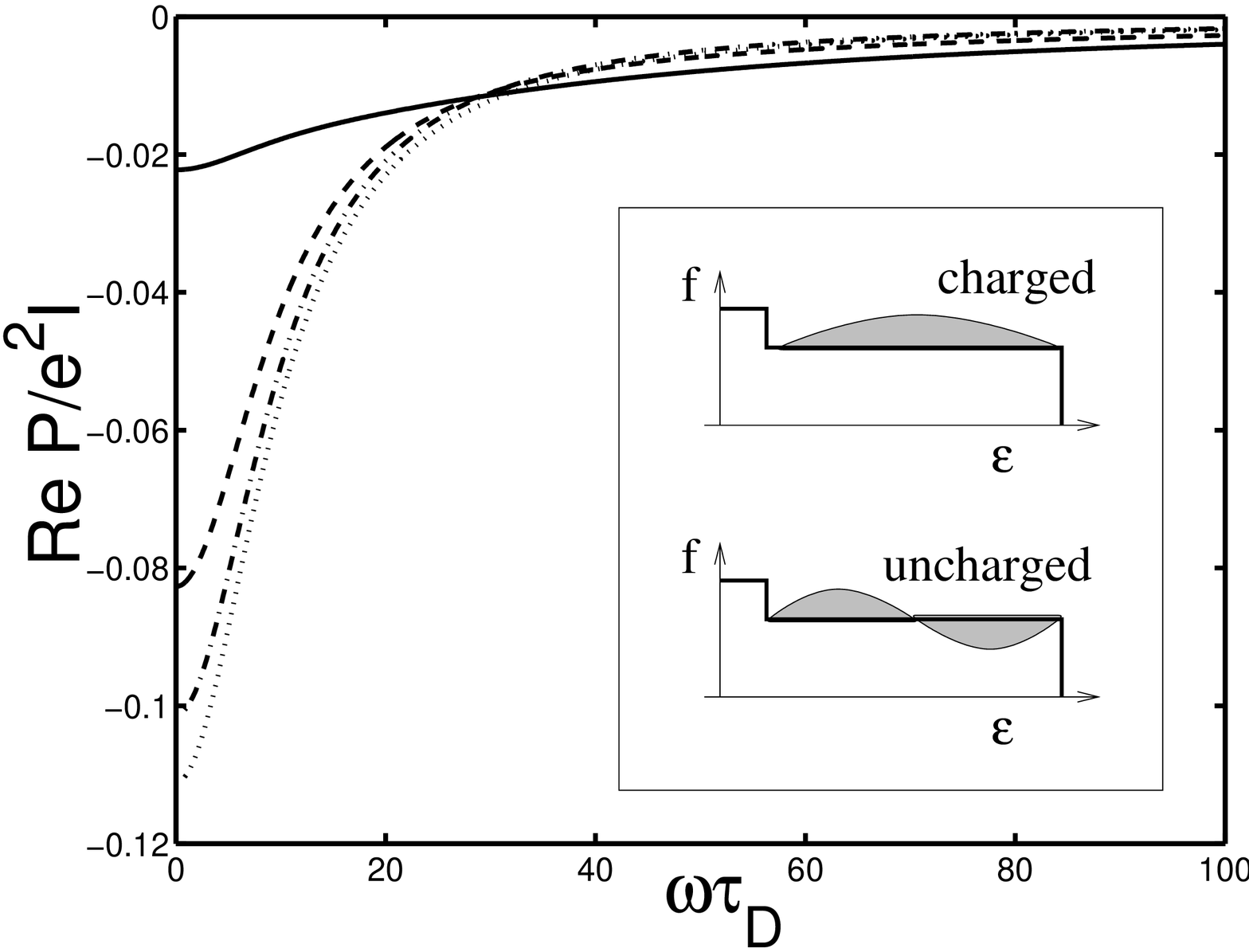}
   \hspace{0.5cm}
   \includegraphics[height=6cm]{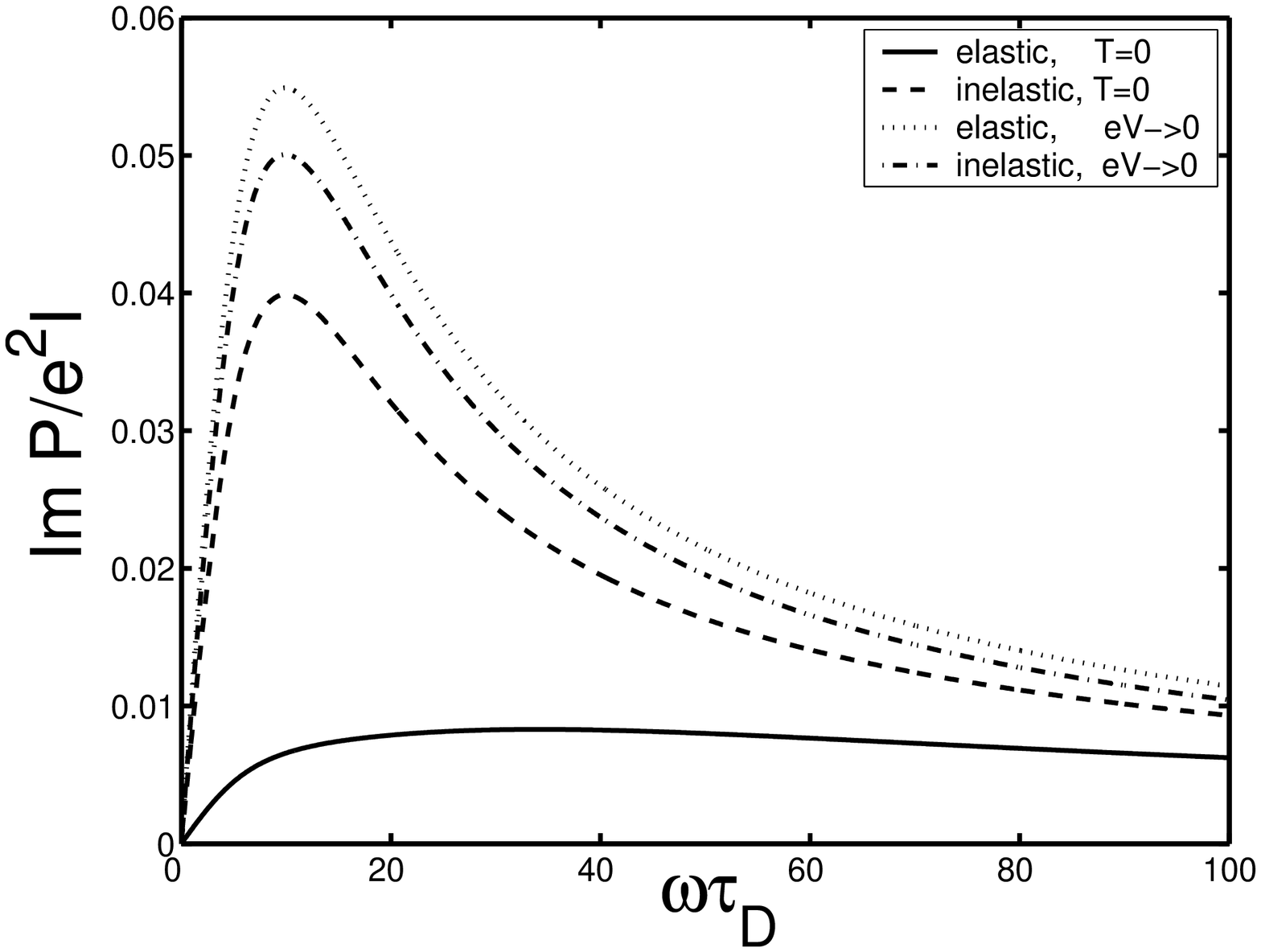}
   \end{tabular}
   \end{center}
   \caption[example]{\label{fig:diff}
  The real and imaginary parts of the ratio $P(\omega)/(e^2I)$
  versus normalized frequency $\omega\tau_D$ ($\tau_D = L^2/D$) for purely
  elastic scattering at $eV \gg T$ (solid lines), hot-electron regime at $eV
  \gg T$ (dashed lines), purely elastic scattering at $eV \ll T$ (dotted lines),
  and hot-electron regime at $eV \ll T$ (dash-dotted lines). Inset: charged and
  uncharged fluctuations of the distribution function $f$. Charged fluctuations
  have a short relaxation time and contribute to fluctuations of current $\delta I$.
  Uncharged fluctuations do not contribute to $\delta I$ directly but affect the
  intensity of noise sources. They decay only via slow diffusion and result in the
  low-frequency dispersion of the third cumulant.}
\end{figure}

In the opposite limit of a strong electron--electron interaction,
the distribution function may be assumed to have a
Fermi shape with a coordinate-dependent temperature $T_e(x)$ and
electric potential $\phi(x)$
\begin{equation}
 f(\eps, x)
 =
 \left[
   1
   +
   \exp
   \left(
    \frac{ \eps - e\phi(x) }{ T_e(x) }
   \right)
 \right]^{-1}
 \label{f-therm}
\end{equation}
and a fluctuation $\delta f$ can be
expressed in terms of fluctuations of these quantities
\begin{equation}
 \delta f(\eps, \br, \omega)
 =
 \frac{
  \partial f(\eps, \br)
 }{
  \partial\phi
 }
 \delta\phi
 +
 \frac{
  \partial f(\eps, \br)
 }{
  \partial T_e
 }
 \delta T_e
 \label{df-hot}
\end{equation}
A substitution of Eq. (\ref{df-hot}) into Eq. (\ref{I^3-gen}) and integration over
the energy readily gives
\begin{equation}
  P_{hot}(\omega)
  =
  \frac{2}{RL}
  \int\limits_{-L/2}^{L/2} dx
  \la
    \delta T_e(x)
    \delta I
  \ra_{\omega}.
  \label{P-hot-2}
\end{equation}
To calculate the correlator in Eq. (\ref{P-hot-2}), we have to
obtain a Langevin-type equation for $\delta T_e$. To this end, we
multiply Eq. (\ref{kinetic}) by $\eps$ and integrate it over
$\eps$, like it was done when deriving the equation of heat
balance\cite{Nagaev-95,Kozub-95}. This gives
\begin{eqnarray}
 \left(
  \frac{\partial}{\partial t}
  -
  D\nabla^2
 \right)
 \left(
  \frac{\pi^3}{3}
  T_e\delta T_e
 \right)
 -
 D\nabla^2
 \left(
  e^2
  \phi\delta\phi
 \right)
 =
 -
 \int d\eps\,\eps
 \nabla\delta{\bf F}^{imp}.
 \label{Langevin-heat}
\end{eqnarray}
By solving this equation together with Eq. (\ref{dphi}) and making use of the
mean effective temperature\cite{Nagaev-95}
\begin{equation}
 \bar{T}_e(x)
 =
 \left[
  T^2
  +
  \frac{3}{\pi^2}
  (eV)^2
  \left(
   \frac{1}{4}
   -
   \frac{x^2}{L^2}
  \right)
 \right]^{1/2},
\end{equation}
one obtains
\begin{equation}
 P_{hot}(\omega)
 =
 -\frac{12}{\pi^2}
 \frac{e^2V}{R}
 \frac{1}{ q_{\omega}^2 L^2 }
 \left[
  1
  -
  \frac{2}{ q_{\omega} L}
  \tanh
  \left(
   \frac{ q_{\omega} L }{2}
  \right)
 \right].
 \label{P_hot-zV}
\end{equation}
at high temperatures $T \gg eV$ and
\begin{eqnarray}
 P_{hot}(\omega)
 =
 \frac{12}{\pi}
 \frac{e^2V}{R}
 \sum\limits_{k=0}^{\infty}
%\nonumber\\
%\times
 \frac{
   J_0(\pi k + \pi/2)
   [
    J_1(\pi k + \pi/2)
    -
    (-1)^k
   ]
 }{
   (2k + 1)
   [
    \pi^2
    (2k + 1)^2
    +
    i\omega
    L^2/D
   ]
 },
\label{series}
\end{eqnarray}
at low temperatures $eV \gg T$, where $J_0$ and $J_1$ are Bessel functions
of order 0 and 1. The corresponding limiting values $P_3(0,0) =
-(3/\pi^2)e^2V/R$ at $eV \ll T$ and $P_3(0,0) = -(8/\pi^2 - 9/16)e^2V/R$
at $eV \gg T$ coincide with the results of Gutman and
Gefen.\cite{Gutman-03}

Figure \ref{fig:diff}  shows that both real and imaginary parts of $P$ have the
most pronounced dispersion in the case of a high temperature or for strong
electron-electron scattering, i.e. when the local distribution function has a
nearly Fermian shape. This unexpected result is in a sharp contrast with the
dispersion of quantum noise,\cite{Altshuler-94} which results from sharp
singularities in the energy dependence of the distribution function. This
difference comes from the different symmetry of the second-order correlation
functions appearing in Eq. (\ref{I^3-gen}) in the elastic case at low
temperatures and in the hot-electron regime. In the former case, the
correlation function $\la\delta f(\eps,x)\delta I\ra_{\omega}$ is an odd
function of the coordinate, but in the latter case $\la\delta T_e(x)\delta
I\ra_{\omega}$ is an even function of $x$.

\subsection{Chaotic Cavity}
\label{sect:chaotic}

  \begin{figure}
   \begin{center}
   \begin{tabular}{c}
   \includegraphics[height=4cm]{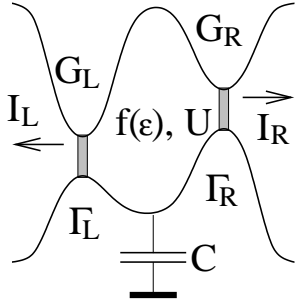}
   \end{tabular}
   \end{center}
   \caption[example]{\label{fig:cavity}
   A chaotic cavity capacitively coupled to a gate}
  \end{figure}

The cascaded Langevin approach may be successfully applied not only to systems
described by Langevin equations with Poissonian statistics of extraneous
sources, which takes place in the case of a Boltzmann kinetics. It can be
equally well applied to systems with arbitrary statistics of noise sources
provided that the conditions of time-scale separation and smallness of the
fluctuations are observed. An example of a semiclassical system with
non-Poissonian sources of noise is a classical chaotic cavity, i.e.  a
metallic island of irregular shape connected to the electrodes $L$, $R$ via two
quantum point contacts of conductances $G_{L,R} \gg e^2/h$ and arbitrary
transparencies $\Gamma_{L,R}$ (see Fig. \ref{fig:cavity}). As the dwell time of
an electron in the cavity $\tau_D = e^2 N_F/(G_L + G_R)$ is much larger than
the time of flight through the cavity, the electrons in the cavity lose memory
of their initial phase and are described by an energy-dependent distribution
function $f(\eps, t)$. Hence the low-frequency dynamics of $f$ is described by
a semiclassical equation, while quantum contact serve as generators of noise
with essentially non-Poissonian statistics. The fluctuations of the electric
current in the left and right contacts are given by equations
\begin{equation}
 \delta I_{L,R}
 =
 \int d\eps\,
 \left[
  (
   \tilde I_{L,R}
  )_{\eps}
  +
  \frac{1}{e}
  G_{L,R}
  \delta f(\eps)
 \right],
 \label{dI}
\end{equation}
where $(\tilde I_L)_{\eps}$ and $(\tilde I_R)_{\eps}$ are the energy-resolved random
extraneous currents generated by the left and right contacts. The fluctuation of the
distribution function $\delta f(\eps)$ obeys a kinetic equation
\begin{equation}
 \left(
  \frac{\partial}{\partial t}
  +
  \frac{1}{\tau_D}
 \right)
 \delta f(\eps,t)
 =
 -
 e
 \frac
 {
  \partial\delta U
 }{
  \partial t
 }
 \frac
 {
  \partial f
 }{
  \partial\eps
 }
 -
 \frac{1}{eN_F}
 [
  (\tilde I_L)_{\eps}
  +
  (\tilde I_R)_{\eps}
 ],
 \label{kinetic}
\end{equation}
where $\tau_D = e^2N_F/(G_L + G_R)$ is the dwell time of an electron in the cavity,
$N_F$ is the density of states in it, and $\delta U$ is a fluctuation of the electric
potential of the cavity. This fluctuation is obtained from the charge-conservation law
\begin{equation}
 \frac
 {
  \partial\delta U
 }{
  \partial t
 }
 =
 \frac{1}{C}
 \frac
 {
  \partial\delta Q
 }{
  \partial t
 }
 =
 -\frac{1}{C}
 \int d\eps\,
 [
  (\tilde I_L)_{\eps}
  +
  (\tilde I_R)_{\eps}
 ],
 \label{charge}
\end{equation}
where $C$ is the electrostatic capacitance of the cavity. Equations (\ref{dI}) -
(\ref{charge}) suggest that different parts of $\delta f$ are described by different
time scales. The relaxation of electrically neutral fluctuations is described by the
characteristic time $\tau_D$, whereas the fluctuations of charge are described by
$\tau_Q = [(G_L + G_R)/C + 1/\tau_D]^{-1}$, which is much shorter than $\tau_D$ for good
conductors.

If the distribution function of electrons in the cavity $f(\eps,t)$ were not allowed to
fluctuate, the contacts would be independent generators of current noise whose
zero-frequency energy-resolved cumulants $\la\la \tilde{I}_{L,R}^n \ra\ra_{\eps}$ could
be obtained from a quantum-mechanical formula
$$
 \la\la
  \tilde{I}_{L,R}^n
 \ra\ra_{\epsilon}
 =
 \frac{G_{L,R}}{\Gamma_{L,R}}
 \frac{\partial^n}{\partial \chi^n}
 \ln
 \bigl\{
   1
   +
   \Gamma_{L,R} f_{L,R}(\epsilon)
   [1-f(\eps)]
   (
    e^{-e\chi}-1)
$$ \begin{equation}
   +
   \Gamma_{L,R}
   f(\eps)
   [1-f_{L,R}(\epsilon)]
   (
    e^{e\chi}-1
   )
 \bigr\}|_{\chi=0}.
 \label{Point Contacts}
\end{equation}
If the voltage is high enough, the noise of isolated contacts can be
considered as white at frequencies at which the distribution function $f$
fluctuates. This allows us to consider the contacts as independent
generators of white noise, whose intensity is determined by the
instantaneous distribution function of electrons in the cavity. Based on
this time-scale separation, we perform a recursive expansion of higher
cumulants of current in terms of its lower cumulants as described in
Section \ref{sect:general}. In the low-frequency limit, the expressions
for the third and fourth cumulants coincide with those obtained by
quantum-mechanical methods for arbitrary ratio of conductances $G_L/G_R$
and transparencies $\Gamma_{L,R}$ \cite{Nagaev-02b}. Very recently,
similar equations for the zero-frequency limit were obtained as a
saddle-point expansion of a stochastic path integral \cite{Pilgram-02}.
  \begin{figure}
   \begin{center}
   \begin{tabular}{c}
   \includegraphics[height=7cm]{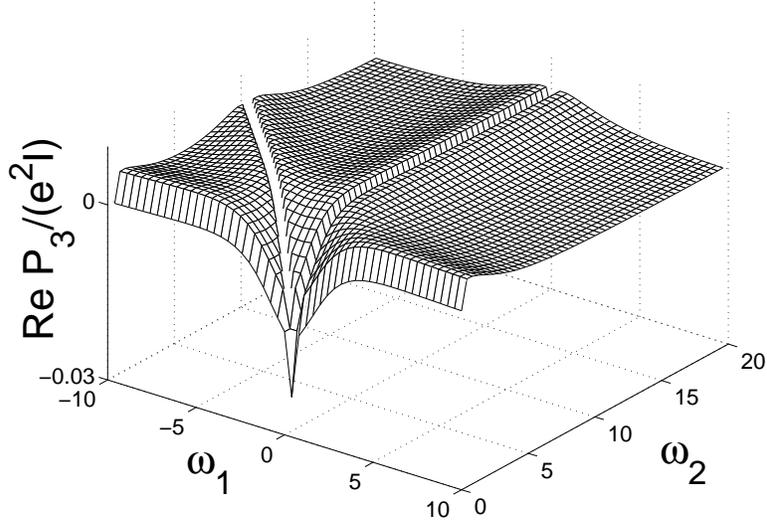}
   \end{tabular}
   \end{center}
   \caption[example]{\label{fig:3D}A 3D plot of ${\rm Re}\,P_3(\omega_1, \omega_2)$ for
   $G_L/G_R=1/2$, $\Gamma_L = \Gamma_R = 3/4$, $\tau_Q = 1/3$, and
   $\tau_D=10$ (dimensionless units). }
  \end{figure}
  \begin{figure}
   \begin{center}
   \begin{tabular}{c}
   \includegraphics[height=7cm]{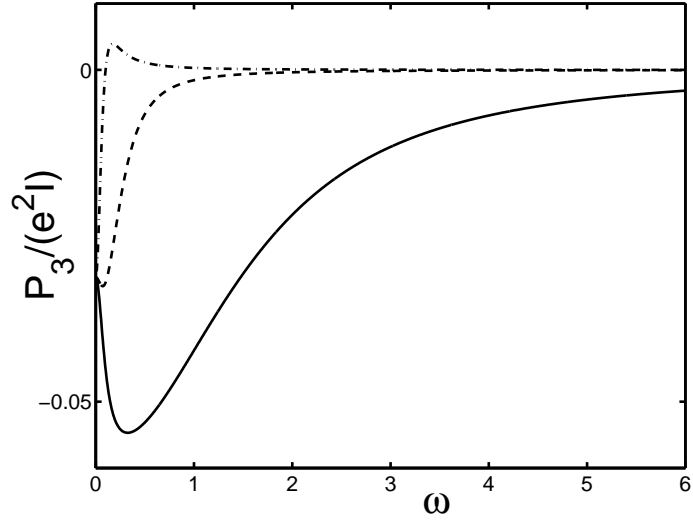}
   \end{tabular}
   \end{center}
   \caption[example]{\label{fig:2D}$P_3(\omega, 0)$ as a
   function of $\omega$ for $G_L/G_R=1$, $\Gamma_L=1$, $\Gamma_R=0$, and
   $\tau_D = 10$ (dimensionless units). The solid, dashed, and dash-dot
   curves correspond to $\tau_Q =1/2$, $\tau_Q =3$, and to weak electrostatic
   coupling $\tau_Q = \tau_D$ ($C=\infty$). }
  \end{figure}
Consider now the frequency dependence of the third cumulant. We will be
interested in the case of a good conductor where the charge-relaxation
time $\tau_Q$ is much shorter than the dwell time $\tau_D$. Unlike the
second cumulant of current, the third cumulant $P_3(\omega_1, \omega_2)$
in general exhibits a strong dispersion at $\omega_{1,2}\sim 1/\tau_D$
\cite{Nagaev-03}. In this limit,
$$
 P_3(\omega_1, \omega_2)
 =
 e^2 I
 \Biggl\{
  3G_L G_R
  \frac
  {
   \left[
    (1 - \Gamma_R) G_L^2
    -
    (1 - \Gamma_L) G_R^2
   \right]^2
  }{
   (G_L + G_R)^6
  }
  -
  2
  \frac
  {
   \Gamma_R^2 G_L^5
   +
   \Gamma_L^2 G_R^5
  }{
   (G_L + G_R)^5
  }
  +
  3
  \frac
  {
   \Gamma_R G_L^4
   +
   \Gamma_L G_R^4
  }{
   (G_L + G_R)^4
  }
  -
  \frac
  {
   G_L^3 + G_R^3
  }{
   (G_L + G_R)^3
  }
$$
\begin{equation}
  -
  G_L G_R
  \frac
  {
   \left[
    (1 - \Gamma_R) G_L^2
    -
    (1 - \Gamma_L) G_R^2
   \right]
   \left(
    \Gamma_R G_L^2
    -
    \Gamma_L G_R^2
   \right)
  }{
   (G_L + G_R)^6
  }
  \left[
   \frac{1}{ 1 + i\omega_1\tau_D }
   +
   \frac{1}{ 1 + i\omega_2\tau_D }
   +
   \frac{1}
   {
    1 - i(\omega_1 + \omega_2)\tau_D
   }
  \right]
 \Biggr\}.
 \label{S-low}
\end{equation}
This dispersion vanishes for symmetric cavities and cavities with two
tunnel or two ballistic contacts. For a cavity with two tunnel contacts,
the white-noise sources (\ref{Point Contacts}) are linear functionals of
the distribution function and hence are not affected by charge-neutral
fluctuations. For the case of two ballistic contacts, Eq. (\ref{Point
Contacts}) depends only on $f$ and not on $f_i$. Then the low-frequency
dispersion does not show up either because fluctuations of current and
distribution function are uncorrelated due to the symmetry of the noise
sources. The shape of $P_3(\omega_1, \omega_2)$ essentially depends on the
parameters of the contacts. In particular, for a cavity with one tunnel
and one ballistic contact with equal conductances $G_L = G_R = G$ it
exhibits a non-monotonic behavior as one goes from $\omega_1 = \omega_2 =
0$ to high frequencies. A relatively simple analytical expression for this
case may be obtained if $\tau_D \gg \tau_Q$ and one of the frequencies is
zero:
\begin{equation}
 P_3(\omega, 0)
 =
 -\frac{1}{32}
 e^2 I
 \frac
 {
  1
  +
  2\tau_D^2
  \omega^2
  +
  \tau_D^2\tau_Q^2
  %(\tau_D - 2\tau_Q)
  \omega^4
 }{
  (1 + \omega^2\tau_D^2)
  (1 + \omega^2\tau_Q^2)^2
 }.
 \label{g10}
\end{equation}
The $P_3(\omega, 0)$ curve shows a clear minimum at $\omega \sim (\tau_D \tau_Q)^{-1/2}$
and the amplitude of its variation tends to $P_3(0, 0)$ as $\tau_Q/\tau_D \to 0$.

\subsection{Effects of External Circuit on Measurements of the Third Cumulant}
\label{sect:circuit}

So far we applied the cascaded method to systems where the noise was governed
by a microscopic distribution function of electrons. Now we give an example
where the noise in a system is determined by a macroscopic quantity.

Recently, Levitov and Reznikov\cite{Levitov-01} predicted that the third cumulant
of noise in a tunnel contact should present a linear function of voltage independent
of temperature. However when Reulet et al.\cite{Reulet-03} tested these predictions
experimentally, they found very poor agreement of their data with theory. The reason
for this description has been explained by Beenakker et al.\cite{Beenakker-03} by the
feedback of the external circuit using the cascade approach.

In experiments of Reulet and co-workers, the actual measurable quantity was the
voltage drop across a macroscopic resistor $R_S$ connected in series with a
mesoscopic contact with a resistance $R_0$, which produced a non-Gaussian noise (see Fig.
\ref{fig:circuit}).
Naively, one could expect that the third cumulant of voltage drop would be just
proportional to the third cumulant of current generated by the contact, but this is
not the case.

Indeed,  consider the macroscopic resistor and the contact as independent
generators of noise currents $\delta I_S$ and $\delta I_0$. The noise of the
macroscopic resistor is Gaussian with the only nonzero second cumulant $\la\la
I_S^2\ra\ra =2T/R_S$ representing its equilibrium noise, and the noise of the
contact has nonzero second and third cumulants $\la\la I_0^2\ra\ra$ and $\la\la
I_0^3\ra\ra$, which depend on the voltage across it. The Kirchhoff's law for
the fluctuation of the total current in the circuit may be written in a form
\begin{equation}
 \delta I
 =
 -\frac{\delta\varphi}{R_S} + \delta I_S
 =
 \frac{\delta\varphi}{R_0} + \delta I_0,
 \label{Kirchhoff}
\end{equation}
where $\delta\varphi$ is the fluctuation of electric potential to be measured. Hence
\begin{equation}
 \delta\phi
 =
 \frac{R_0 R_S}{R_0 + R_S}
 ( \delta I_S - \delta I_0).
 \label{dphi-sol}
\end{equation}
One obtains the second cumulant of voltage just by multiplying and averaging
two solutions (\ref{dphi-sol}), which gives
\begin{equation}
 \la\la
  \varphi^2
 \ra\ra
 =
 \left(
  \frac{R_0 R_S}{R_0 + R_S}
 \right)^2
 \left(
  \la\la I_0^2 \ra\ra
  +
  2\frac{T}{R_S}
 \right).
 \label{dphi^2}
\end{equation}
Calculations of the third cumulant of voltage according to Eq. (\ref{xxx-symb}) give
\begin{equation}
 \la\la \varphi^3 \ra\ra
 =
 -
 \left(
  \frac{R_0 R_S}{R_0 + R_S}
 \right)^3
 \la\la I_0^3 \ra\ra
 +
 3
 \left(
  \frac{R_0 R_S}{R_0 + R_S}
 \right)^4
 \left(
  2\frac{T}{R_S}
  +
  \la\la I_0^2 \ra\ra
 \right)
 \frac{ d\la\la I_0^2 \ra\ra }{d\varphi}.
 \label{dphi^3}
\end{equation}
The first term in this expression is the naively expected result, and the second term
presents the cascade correction that results from modulations of the second cumulant of
current generated by the mesoscopic contact by fluctuations of voltage across it.
Equation (\ref{dphi^3}) suggests that the feedback of the external circuit is very
essential even at $R_S \ll R_0$. For a tunnel contact where
$\la\la I_0^2\ra\ra = (e\varphi/R_0)\coth(e\varphi/2T)$ and $\la\la I_0^3\ra\ra = e^2\varphi/R_0$, it results
in a change of slope of $\la\la \varphi^3 \ra\ra$ versus $V$ curve from negative at
$eV \ll T$ to positive at $eV \gg T$. This is exactly what is observed in experiment.

  \begin{figure}
   \begin{center}
   \begin{tabular}{c}
   \includegraphics[width=7cm]{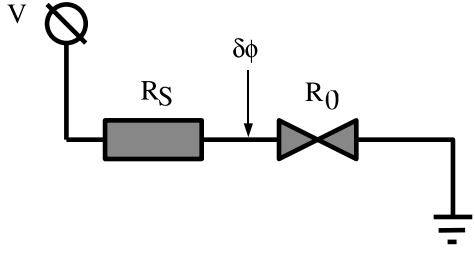}
   \end{tabular}
   \end{center}
   \caption[example]{\label{fig:circuit} A mesoscopic contact in series with a
   macroscopic resistor. The cumulants of current of the contact depend on the
   fluctuations of potential $\delta\phi$. }
  \end{figure}

\section{Summary}
\label{summary}

In summary, we demonstrated a semiclassical method for calculating higher
cumulants of noise in systems where a large separation exists between the
characteristic times describing the correlations of microscopic sources of
noise and the characteristic times describing the dynamics of averages. The
calculations of the frequency-dependent third cumulants in a number of
mesoscopic systems show that their dispersion may be very different from that
of conventional noise and is very sensitive to fine properties of these systems
that almost do not manifest themselves otherwise. As the conditions for
validity of the method are rather general, it may be applied to macroscopic
systems, too.

\end{document}